\documentclass[11pt,a4paper]{article}
\usepackage{jheppub}
\usepackage[T1]{fontenc}
\usepackage{amssymb}
\usepackage{mathtools}
\usepackage{xcolor}

\usepackage{hyperref}
\usepackage{epsfig}
\usepackage{amsmath,latexsym,amssymb}
\usepackage{graphicx}
\usepackage{braket}
\usepackage{ulem}
\usepackage[noabbrev]{cleveref}

\usepackage{mathrsfs}
\def\be{\begin{equation}}
\def\ee{\end{equation}}
\def\ba{\begin{eqnarray}}
\def\ea{\end{eqnarray}}

\newcommand{\la}{\lambda}

\newcommand{\bh}{\bar{h}}

\newcommand{\bx}{\bar{x}}

\newcommand{\zbar}{\bar{z}}

\newcommand{\hy}{{}_{2}F_1}

\def\D{\Delta}

\newcommand{\comment}[1]{}

\def\fc#1#2{{\frac{#1}{#2}}}

\newcommand{\eea}{\end{eqnarray}}

\author{
Iustin Surubaru${}^{1}$,\, Bin Zhu${}^{1}$\\[0.5cm]

$^1${\it School of Mathematics and Maxwell Institute for Mathematical Sciences,\\ University of Edinburgh,
EH9 3FD, U.K. }\\[0.2cm]
}

\emailAdd{iustin.surubaru@ed.ac.uk}
\emailAdd{bzhu@ed.ac.uk}

\title{Conformal blocks from celestial graviton amplitudes}

\abstract{Four-point gluon and graviton correlators in celestial holography are famously non-analytic, having distributional support. In this work, we propose an alternative graviton correlator that is analytic and displays several desirable properties. We compute the four-point correlator involving one graviton shadow operator and three graviton primary operators from the celestial four-point graviton amplitudes at tree-level. We perform the conformal block decomposition for the shadow correlator in the compatible channel. For the case when the shadow operator is conformally soft, we compute the single-valued completion of the shadow correlator and perform the conformal block decomposition of the single-valued shadow correlator in all channels. We find an integral representation of the single-valued shadow correlator, which allows us to invert the shadow transform to find the single-valued celestial graviton amplitude. We study various properties of the single-valued celestial graviton amplitude. Interestingly, it exhibits a double copy structure in relation to its counterpart gluon amplitude.}

\gdef\@fpheader{}
\makeatother

\begin{document}
\maketitle

\section{Introduction}
Celestial amplitudes for massless particles obtained from expressing scattering
amplitudes in  boost-eigenstate basis \cite{Pasterski:2016qvg,Pasterski:2017kqt,Pasterski:2017ylz,Stieberger:2018onx,Arkani-Hamed:2020gyp} have proven very useful in making the conformal properties of scattering amplitudes manifest \cite{Pasterski:2016qvg,Pasterski:2017kqt,Pasterski:2017ylz,Stieberger:2018onx,Arkani-Hamed:2020gyp,Donnay:2018neh,Fan:2019emx,Pate:2019mfs,Adamo:2019ipt,Guevara:2019ypd,Puhm:2019zbl,Pate:2019lpp,Fotopoulos:2019vac,Fotopoulos:2020bqj,Banerjee:2020kaa,Donnay:2020guq,Guevara:2021abz,Strominger:2021mtt,Himwich:2021dau}. By construction, celestial amplitudes transform under $\mathrm{SL}(2, \mathbb{C})$ Lorentz transformations as two-dimensional CFT correlators of primary fields. The main proposal of celestial holography is that celestial amplitudes are correlators of the putative two-dimensional celestial CFT (CCFT) that holographically describes quantum gravity in four-dimensional asymptotically flat space \cite{Strominger:2017zoo,Pasterski:2021rjz,Raclariu:2021zjz,Donnay:2023mrd}. However, due to the kinematic constraints of translation invariance in the bulk, low-point celestial amplitudes are distributional in the coordinates on the
celestial sphere \cite{Pasterski:2017ylz}. Therefore, CCFT, if it exists, is not a CFT of garden variety.

There have been different methods of resolving the distributional nature of celestial amplitudes, including breaking the translation symmetry in the bulk by considering backgrounds \cite{Fan:2022vbz,Casali:2022fro,Fan:2022kpp,deGioia:2022fcn,Gonzo:2022tjm,Pasterski:2020pdk,Banerjee:2023rni,Ball:2023ukj,Crawley:2023brz,Adamo:2024mqn, Costello:2022wso,Melton:2022fsf,Bittleston:2023bzp,Costello:2022jpg,Costello:2023hmi,Adamo:2023zeh,Melton:2023dee,Lipstein:2023pih,Taylor:2023ajd,Bittleston:2024rqe,Adamo:2024xpc,Tropper:2024evi,McLoughlin:2024ldp,Bogna:2024gnt,Stieberger:2022zyk,Taylor:2023bzj,Stieberger:2023fju}, computing celestial leaf amplitudes, which can be combined into the distributional celestial amplitudes \cite{Melton:2023bjw,Melton:2024akx,Duary:2024cqb,Mandal:2024zao} and through a change of basis \cite{Fan:2021isc,Fan:2021pbp,Sharma:2021gcz,Kapec:2021eug,Hu:2022syq,Banerjee:2022hgc,Kapec:2022axw,De:2022gjn,Chang:2022jut,Jorge-Diaz:2022dmy,Fan:2023lky,Furugori:2023hgv,Narayanan:2024qgb, Banerjee:2024hvb,Guevara:2021tvr,Kulp:2024scx}. A particular change of basis, the {\itshape shadow conformal basis}, obtained from the shadow transform of the boost eigenstate basis, has been particularly useful in celestial holography. It has been used in the study of state-operator correspondence in CCFT \cite{Crawley:2021ivb}, the study of vacuum entanglement entropy in free Maxwell theory
in four-dimensional Minkowski spacetime \cite{Chen:2023tvj,Chen:2024kuq}, and the shadow transform of the conformally soft graviton operator has been used in constructing the stress tensor of the CCFT \cite{Kapec:2014opa,Kapec:2016jld}. Despite the importance of the shadow conformal basis, the celestial correlator involving one graviton shadow operator and three graviton primary operators has not been computed. In this paper, we perform the computation and find that the resulting object has many interesting properties.

 The correlator involving one gluon shadow operator and three gluon primary operators was computed in \cite{Fan:2021isc}. In \cite{Fan:2021pbp}, the single-valued completion of the shadow correlator was constructed in the conformally soft gluon limit. Interestingly, the single-valued shadow correlator of gluons can be expressed in terms of a complex integral reminiscent of the Coulomb gas formalism in minimal models \cite{Dotsenko:1984ad,Dotsenko:1984nm}. This integral representation was used to find the inverse shadow which results in single-valued celestial gluon amplitudes. In this work, we apply the methods in \cite{Fan:2021isc,Fan:2021pbp} to celestial graviton amplitudes. The conformal partial wave expansion or conformal block decomposition has been a useful tool in the study of celestial amplitudes (see, e.g. \cite{Fan:2021isc,Fan:2021pbp,Lam:2017ofc,Chang:2021wvv,Nandan:2019jas,Law:2020xcf,Atanasov:2021cje,Chang:2022jut}). We perform conformal block decompositions of various correlators computed in this work and investigate their properties.

 The paper is organized as follows. In section \ref{sec2}, we start from the celestial MHV 4-point graviton amplitude and compute its shadow transform. We perform conformal block decompositions of the shadow correlator in the compatible channel. In section \ref{sec3}, we compute the single-valued completion of the shadow correlator of graviton when the shadow field is conformally soft. We obtain conformal block decompositions of the single-valued shadow correlator in all three channels. We extract the leading OPEs from the single-valued shadow correlator and compare them with the known celestial OPEs. We find an integral representation of the single-valued shadow correlator. In section \ref{sec4}, utilizing the integral representation, we invert the shadow transform and find the single-valued celestial graviton amplitude and study its basic properties. We show that it has a double copy structure compared to its counterpart from gluon amplitudes. In section \ref{sec5}, we discuss briefly some future directions.

\section{Shadow transform of the four-graviton celestial amplitudes} \label{sec2}
The celestial MHV 4-point graviton amplitude at tree-level is given by \cite{Stieberger:2018edy,Puhm:2019zbl}
\begin{equation}
\begin{split}
    &\langle \phi_{\Delta_1,-}(z_1,\Bar{z}_1)\phi_{\Delta_2,-}(z_2,\Bar{z}_2)\phi_{\Delta_3,+}(z_3,\Bar{z}_3)\phi_{\Delta_4,+}(z_4,\Bar{z}_4)\rangle \\
    =& z^{10/3}(1-z)^{-2/3}\delta(z-\Bar{z})\delta\left(\sum_{i=1}^4\lambda_i-2i\right) \prod_{i<j}z_{ij}^{h/3-h_i-h_j}\Bar{z}_{ij}^{\Bar{h}/3-\Bar{h}_i-\Bar{h}_j}\, , \label{eq:4graviton}
\end{split}
\end{equation}
where $\phi_{\Delta_i,\pm}(z_i,\bar{z}_i)$ is a conformal primary graviton of dimension $\Delta_i$ and helicity $\pm2$ inserted at position $(z_i,\bar{z}_i)$ on the celestial sphere with the conformal cross ratios
\begin{equation}
    z=\frac{z_{12}z_{34}}{z_{13}z_{24}} \,  , \quad \bar{z}=\frac{\bar{z}_{12}\bar{z}_{34}}{\bar{z}_{13}\bar{z}_{24}} .
\end{equation}
The delta function of the cross ratios $\delta(z-\bar{z})$ enforces planarity of four-particle scattering. 
One has to specify the incoming and outgoing configuration for a given two-two scattering process, which imposes a constraint on the cross ratio,
\begin{equation}
    \begin{split}
        &a) \,12 \rightleftharpoons 34\, , \quad z>1 \, ,\\
        &b) \, 13 \rightleftharpoons 24\, , \quad  0<z<1 \, ,\\
        &c) \, 14 \rightleftharpoons 23\, , \quad z<0  \, .
    \end{split}
\end{equation}
The conformal dimensions of the two positive helicity gravitons are taken to be in the principle series, $\Delta_3=1+i\lambda_3$, $\Delta_4= 1+i\lambda_4$ \cite{Pasterski:2017kqt} \footnote{The principle series is chosen such that the conformal wavefunctions are $\delta$-function normalizable with respect to the standard inner product. One can go beyond the principle series as shown in \cite{Donnay:2020guq}.}. The conformal dimensions of the two negative helicity gravitons are taken to be $\Delta_1=i\lambda_1'=1+i\lambda_1$, $\Delta_2=i\lambda_2'=1+i\lambda_2$ so that the celestial amplitude (\ref{eq:4graviton}) is finite with the constraint imposed by $\sum_{i=1}^4\lambda_i=2i$. The helicity of the graviton is related to the conformal weights by $J=h-\bar{h}$, where
\begin{equation}
\begin{split}
    h_1=-\frac{1}{2}+\frac{i\lambda_1}{2},\quad &\Bar{h}_1=\frac{3}{2}+\frac{i\lambda_1}{2} \, ,\\  
    h_2=-\frac{1}{2}+\frac{i\lambda_2}{2},\quad&\Bar{h}_2=\frac{3}{2}+\frac{i\lambda_2}{2} \, ,\\
    h_3=\frac{3}{2}+\frac{i\lambda_3}{2},\quad&\Bar{h}_3=-\frac{1}{2}+\frac{i\lambda_3}{2} \, ,\\
    h_4=\frac{3}{2}+\frac{i\lambda_4}{2},\quad&\Bar{h}_4=-\frac{1}{2}+\frac{i\lambda_4}{2} \, .
\end{split}
\end{equation}
The distributional nature of the low-point celestial amplitudes makes it difficult to apply standard CFT techniques to celestial CFT.

To resolve this issue, we follow the method in \cite{Fan:2021isc} by computing the shadow correlator\footnote{The shadow transform of a primary operator is defined by $\tilde{\phi}(w,\bar{w}) = \int d^2 y(w-y)^{2h-2} (\bar{w}-\bar{y})^{2\bar{h}-2} \phi(y,\bar{y})$, modulo an overall constant that we will omit. The shadow operator has conformal weights $1-h, 1-\bar{h}$, hence with the conformal dimension $\tilde{\Delta}=2-\Delta$, and helicity $\tilde{J}=-J$.}:
\begin{equation}
\begin{split}
&\langle \tilde{\phi}_{\tilde{\Delta}_1,+}(z_{1'},\Bar{z}_{1'})\phi_{\Delta_2,-}(z_2,\Bar{z}_2)\phi_{\Delta_4,+}(z_3,\Bar{z}_3)\phi_{\Delta_3,+}(z_4,\Bar{z}_4)\rangle \\
   =& \int\frac{d^2 z_1}{(z_1'-z_1)^{3-i\lambda_1}(\Bar{z}_1'-\Bar{z}_1)^{-1-i\lambda_1}} \langle \phi_{\Delta_1,-}(z_1,\Bar{z}_1)\phi_{\Delta_2,-}(z_2,\Bar{z}_2)\phi_{\Delta_4,+}(z_3,\Bar{z}_3)\phi_{\Delta_3,+}(z_4,\Bar{z}_4)\rangle  \, . \label{eq:defshadow}
\end{split}
\end{equation}
The cross ratios for the shadow correlator are defined as
\begin{equation}
    z'=\frac{z_{1'2}z_{34}}{z_{1'3}z_{24}} \, , \quad \bar{z}'=\frac{\bar{z}_{1'2}\bar{z}_{34}}{\bar{z}_{1'3}\bar{z}_{24}} \, .
\end{equation}
Note that the shadow operator has conformal weights $(h_{1'},\bh_{1'})=(\frac{3}{2}-\frac{i\lambda_1}{2},-\frac{1}{2}-\frac{i\lambda_1}{2})$. To perform the shadow transform in (\ref{eq:defshadow}), we change the integration variable to $y=\frac{z}{z'}$. The corresponding Jacobian is $|dz_1/dy|^2$, where
\begin{equation}
    \frac{dz_1}{dy}=\kappa^2 z_{23}z_{1'2}z_{1'3}\,  ,  \quad   \kappa =\frac{z_{13}}{z_{1'3}z_{23}}=( y\, z_{21'} +z_{1'3})^{-1}  \, .
\end{equation}
All the  $z_{ij}$ involving $z_1$ can be expressed in terms of the new variables as:
\begin{equation}
    z_{12}=\kappa y z_{1'2}z_{23} \, ,\quad z_{13}=\kappa z_{23}z_{1'3} \, ,\quad z_{14}=\kappa\left(y-\frac{1}{z'}\right)z_{34}z_{1'2},\quad z_{11'}=\kappa(y-1)z_{1'2}z_{1'3} \, .
\end{equation}
Combining (\ref{eq:4graviton}) with (\ref{eq:defshadow}), we find
\begin{equation}
\begin{split}
    &\langle \tilde{\phi}_{\tilde{\Delta}_1,+}(z_{1'},\Bar{z}_{1'})\phi_{\Delta_2,-}(z_2,\Bar{z}_2)\phi_{\Delta_4,+}(z_3,\Bar{z}_3)\phi_{\Delta_3,+}(z_4,\Bar{z}_4)\rangle \\
    =&(z_{1'2}\bar{z}_{1'2})^{\frac{i(\lambda_1-\lambda_2)}{2}}(z_{1'3}\bar{z}_{1'3})^{-\frac{i(\lambda_3+\lambda_4)}{2}}(z_{23}\bar{z}_{23})^{2+i\lambda_4}(z_{24}\bar{z}_{24})^{1+\frac{i(\lambda_3-\lambda_4)}{2}}(z_{34}\bar{z}_{34})^{1+\frac{i(\lambda_1+\lambda_2)}{2}}\\
    &\times \frac{z_{23}}{z_{1'2}z_{1'3}^3z_{24}z_{34}^3}\frac{\bar{z}_{1'2}^3\bar{z}_{34}^5}{\bar{z}_{1'3}^3\bar{z}_{23}^3\bar{z}_{24}^5} I_s(z',\bar{z}') \, , \label{eq:4graviton1shadow}
\end{split}
\end{equation}
where the $y$ integral reads:
\begin{equation}
    I_s(z',\bar{z}')=\int d^2 y \delta(z'y-\bar{z}'\bar{y})\frac{(z'y-1)^{-2-i(\lambda_1+\lambda_4)}}{(yz-1)^{3-i\lambda_1}(\bar{y}\bar{z}'-1)^{-1-i\lambda_1}}|y|^{2-i(\lambda_1+\lambda_2)} \theta(|z'y|-1)\, .
\end{equation}
where we have chosen the  four-dimensional scattering channel to be $\,12 \rightleftharpoons 34$ by including the Heaviside step function $\theta(|z'y|-1)$ in the integrand. After some manipulations, the $y$ integral can be written as
\begin{equation}
    I_s(z', \bar{z}')=\int_1^\infty \frac{dr}{(z'\bar{z}')^{2-\frac{i(\lambda_1+\lambda_2)}{2}}}\frac{(r-1)^{-2-i\lambda_4}r^{2-i(\lambda_1+\lambda_2)}}{\left(\frac{r}{z'}-1\right)^{3-i\lambda_1}\left(\frac{r}{\bar{z}'}-1\right)^{-1-i\lambda_1}} \, ,
    \label{pluschannelbeforesoft}
\end{equation}
where $r=|z'y|$. We will study the shadow correlator (\ref{eq:4graviton1shadow}) in details in the rest of this section.

\subsection{Conformal block decomposition}
\label{sec:confblockdecomp}
The shadow correlator that we obtained in  (\ref{eq:4graviton1shadow}) is defined on the entire complex plane without any constraints on the cross ratios. As a result, we can decompose it into a sum over conformal blocks for all three two-dimensional channels. In the subsequent sections we will often denote channels using the notation $(ij\rightleftharpoons kl)_{d}$, where the first two indices $i,j$ denote the incoming particles, $k,l$ are the outgoing ones and $d$ is the space-time dimension, which in our case will always be $d=2$. 

We begin with the conformal decomposition of the four-point shadow correlator (\ref{eq:4graviton1shadow}) in the two-dimensional $s$ channel, $(12\rightleftharpoons 34)_{{\mathfrak{2}}}$. This was referred as the `compatible channel' in \cite{Fan:2021isc}. To that end, we
set $z'=x$, $\zbar'=\bar x$. For this channel, we expand correlators around $x=0$. We define \cite{DiFrancesco:1997nk}:
\be
G_{34}^{21}(x,\bar{x}) =  \lim_{z_{1'}, \bar{z}_{1'}\rightarrow \infty} z_{1'}^{\,2 h_{1'}} \bar{z}_{1'}^{\,2 \bar{h}_{1'}}
\Big\langle\tilde\phi_{\tilde\D_1,+}(z_{1'},\zbar_{1'})\phi_{\D_2,-}(1,1)\phi_{\D_3,+}
(z'=x,\zbar'=\bar x)
\phi_{\D_4,+}(0,0)\Big\rangle\ .
\label{eq:G2134}
\ee
We obtain
\begin{equation}
    \begin{split}
       G_{34}^{21}(x,\bx)=   (1-x)^{3+i\lambda_4}x^{-2+\frac{i(\lambda_1+\lambda_2)}{2}}(1-\bar{x})^{-1+i\lambda_4}\bar{x}^{5-\frac{i(\lambda_3+\lambda_4)}{2}}I_s(x,\bar{x}) \, , \label{gg12}
    \end{split}
\end{equation}
with $I_s(x,\bar{x})$ given by (\ref{pluschannelbeforesoft}). In general, $I_s(x,\bar{x})$ can be expressed as an Appell $F_1$ function. See e.g. \cite{Fan:2021isc} for the gluon case. Here, we consider the simplest case where $\lambda_1=i$ \footnote{The corresponding operator has conformal weights $(h,\bar{h})=(-1,1)$. The shadow operator was proposed as the stress tensor in CCFT \cite{Fotopoulos:2019tpe}.}. As we will demonstrate in the following sections, a series of interesting consequences emerges from this simplest case.

In the limit $\lambda_1=i$, there is no $\bar{x}$ contribution inside the integral (\ref{pluschannelbeforesoft}) and the result gets simplified 
\begin{equation}
\begin{split}
    I_s(x,\bar{x})_{\lambda_1=i} &= x^{\frac{3}{2}+\frac{i\lambda_2}{2}} \bar{x}^{-\frac{5}{2} +\frac{i\lambda_2}{2}} \int_1^\infty dr (r-1)^{-2-i\lambda_4}\ r^{3-i\lambda_2}\ \left(\frac{r}{x}-1\right)^{-4} \\
    &=x^{\frac{3}{2}+\frac{i\lambda_2}{2}} \bar{x}^{-\frac{5}{2}+\frac{i\lambda_2}{2}} B(-i\lambda_3,-i\lambda_4)\,\hy\left({4,-i\la_3\atop 1+i\la_2};x\right)
    ,
\end{split}
\end{equation}
where $B(x,y)$ is the Euler beta function and $\hy\left({a,b\atop c};x\right)$ is the hypergeometric function. Therefore, (\ref{gg12}) in this case becomes
\begin{equation}
    G_{34}^{21}(x,\bx)_{\lambda_1=i}=S_1(x) \bar{I}_1(\bar{x}) \, , \label{eq:GS1I1bar}
\end{equation}
where
\begin{equation}
    S_1(x) = B(-i\lambda_3,-i\lambda_4) (1-x)^{3+i\lambda_4}x^{i\lambda_2-1} \, \hy\left({4,-i\la_3\atop 1+i\la_2};x\right) ,\label{defschan}
\end{equation}
\begin{equation}
    \bar{I}_1(\bar{x})= (1-\bar{x})^{-1+i\lambda_4}\bar{x}^{3+i\lambda_2}. \label{nonholschan}
\end{equation}
A $\mathrm{SL}(2,\mathbb{C})$ conformal block of a primary field with chiral weights $(h,\bar h)$ has the form \cite{Osborn:2012vt}:
\be K_{34}^{21}[h,\bh](x,\bar{x})=\bx^{\bh-\bh_3-\bh_4}\ \hy\left({\bh-\bh_{12},\bh+\bh_{34}\atop 2\bh};\bx\right)x^{h-h_3-h_4}\ \hy\left({h-h_{12},h+h_{34}\atop 2h};x\right)\ .\label{bls}\ee
In the shadow correlator (\ref{eq:4graviton1shadow})
\begin{align}\nonumber&
h_{12}=h_{1'}-h_2= \textstyle\frac{5}{2}- \frac{i\lambda_2}{2}\ , & \bh_{12}=\bh_{1'}-\bh_2=\textstyle-\frac{3}{2}-\frac{i\lambda_2}{2}\ ,\\[1mm] &
h_{34}=h_3-h_4=\textstyle\frac{i\lambda_3}{2}-\textstyle\frac{i\lambda_4}{2}
\ , & \bh_{34}=\bh_3-\bh_4=\textstyle\frac{i\lambda_3}{2}-\textstyle\frac{i\lambda_4}{2}\ ,\\[1mm] &
h_3+h_4=3+\textstyle\frac{i\lambda_3}{2}+\textstyle\frac{i\lambda_4}{2}=\frac{5}{2}-\textstyle\frac{i\lambda_2}{2}\ , & ~~~~~~\bh_3+\bh_4=-1+\textstyle\frac{i\lambda_3}{2}+\textstyle\frac{i\lambda_4}{2}=-\frac{3}{2}-\textstyle\frac{i\lambda_2}{2}\ ,\nonumber
\end{align}
where we used $\lambda_2+\lambda_3+\lambda_4=i$. The antiholomorphic part (\ref{nonholschan}) can be written as a single block corresponding to $\bar{h}=\frac{3}{2}+\frac{i\lambda_2}{2}$,
\begin{equation}
\begin{split}
    \bar{I}_1(\bar{x})&= (1-\bar{x})^{-1+i\lambda_4}\bar{x}^{3+i\lambda_2} = \bar{x}^{3+i\lambda_2}\ \hy\left({3+i\la_2,1-i\la_4\atop 3+i\la_2};\bx\right)\\
    & =\bar{x}^{\bar{h}-\bar{h}_3-\bar{h}_4}\ \hy\left({\bh-\bh_{12},\bh+\bh_{34}\atop 2\bh};\bx\right)
    \, . \label{eq:antisingleblock}
\end{split}
\end{equation}
The holomorphic part (\ref{defschan}) cannot be expressed as a single block. As shown in \cite{Fan:2021isc} for the gluon case, it is straightforward to use the following identities of hypergeometric functions to express the holomorphic part as a sum of conformal blocks: 
\begin{equation}\label{hy1}
\hy\left({a,b\atop c};x\right) = (1-x)^{c-a-b}\ \hy\left({c-a,c-b\atop c};x\right) \, ,\ee
\be \hy\left({a,b\atop c-1};x\right)=\sum_{m=0}^{\infty} \frac{(a)_m (b)_m}{(c-1)_{2m}} x^m \,\hy\left({a+m,b+m\atop c+2m};x\right)\ ,\label{hy2}
\ee
\be
 \hy\left({a,b\atop c};x\right)= \sum_{m=0}^{\infty}  \frac{(-1)^m (a)_m (c-b)_m}{(c)_{2m}} x^m {}_2F_1\left({a+m,b+m+1\atop c+2m+1};x\right) \, ,\label{hy3}\ee 
where $(a)_m=\Gamma(a+m)/\Gamma(a)$ are the Pochhammer symbols.
The holomorphic part (\ref{defschan}) becomes
\begin{equation}
\begin{split}
     S_1(x) &=\sum_{k=1}^\infty x^{k-2+i\lambda_2}a_k\, \hy\left({k-2+i\la_2,k-i\la_4\atop1+2k+i\la_2};x\right)\\
     &=\sum_{k=1}^{\infty}x^{h_k-h_3-h_4}\ a_{k}\, \hy\left({h_k-h_{12},h_k+h_{34}\atop 2h_k};x\right)\, ,
\end{split}
\end{equation}
where
\begin{equation}
    h_k= k+\frac{1}{2}+\frac{i\lambda_2}{2} \, ,
\end{equation}
and the coefficients are
\begin{equation}
    a_k=\frac{(2+k)!}{6}\left(\frac{\Gamma(k-i\lambda_3)\Gamma(-i\lambda_4)}{\Gamma(2k+i\lambda_2)}+(-1)^{k-1}\frac{\Gamma(k-i\lambda_4)\Gamma(-i\lambda_3)}{\Gamma(2k+i\lambda_2)}\right) \, . \label{ak_coeff}
\end{equation}
To summarize, conformal block decomposition of the shadow correlator (\ref{eq:GS1I1bar}) into the two-dimensional $s$ channel is given by
\begin{equation}
     G_{34}^{21}(x,\bx)_{\lambda_1=i}=\sum_{k=1}^{\infty}a_{k} K^{21}_{34}\left[ k+\frac{1}{2}+\frac{i\lambda_2}{2}, \, \frac{3}{2}+\frac{i\lambda_2}{2}\right](x,\bar{x}) \, ,
\end{equation}
where the primary fields have dimensions $\Delta=3+J+i\lambda_2 $, where $J \geq 0$ is an integer spin.

One can also decompose the shadow correlator (\ref{eq:GS1I1bar}) into the other two-dimensional channels: $x\approx 1$, corresponding to the $(14\rightleftharpoons 32)_{{\mathfrak{2}}}$ channel
and $x\approx \infty$, corresponding to the $(13\rightleftharpoons 42)_{{\mathfrak{2}}}$ channel, after analytic continuations. However, similar to the gluon case \cite{Fan:2021isc,Fan:2021pbp}, the shadow correlator is not single-valued upon analytic continuation. This can be fixed by computing its single-valued completion as we will describe in section \ref{sec3}. Moreover, as a result of the analytic continuation, the spectrum contains states with continuous spin depending on $i\la_2$. We will see that these imaginary spin states disappear from the spectrum of the single-valued correlator.

\subsection{Conformal block decomposition for general $\la_1$}
While highlighting important properties about the shadow transformed celestial correlator, the block decomposition in the limit that one of the operators becomes soft is simplified. However, for the $(12\rightleftharpoons 34)_{\mathfrak{2}}$ channel, we can perform a general conformal block decomposition. The greatest advantage of the $\la_1=i$ case is that the shadow integral can be expressed in terms of a hypergeometric function in the holomorhic variables. For general $\la_1$, following \cite{Fan:2021isc}, the correlator can be expressed in terms of the Appell hypergeometric function:
\be
F_1\left({a;b_1,b_2\atop c}; x,y\right)=\fc{\Gamma(c)}{\Gamma(a)\Gamma(c-a)}\int_0^1 \mathrm{d}t\ t^{a-1}(1-t)^{c-a-1}(1-tx)^{-b_1}(1-ty)^{-b_2}.
\ee
With this integral representation, the general (\ref{pluschannelbeforesoft}) integral becomes:
\begin{align}
    I_s(x,\bar{x})=&\frac{(x\bar{x})^{1+\frac{i(\lambda_2-\lambda_1)}{2}}}{(\bar{x})^4}\int_0^1 \mathrm{d}r\ r^{i(\lambda_4+\lambda_2)}(1-r)^{-2-i(\lambda_1+\lambda_4)}(1-rx)^{-3+i\lambda_1}(1-r\bar{x})^{1+i\lambda_1}=\nonumber\\
    =&\frac{(x\bar{x})^{1+\frac{i(\lambda_2-\lambda_1)}{2}}}{(\bar{x})^4} B(1+i\lambda_2+i\lambda_4,1+i\lambda_2+i\lambda_3)\nonumber\\
&\times F_1\left({1+i\lambda_2+i\lambda_4;3-i\lambda_1, -1-i\lambda_1\atop i(\lambda_2-\lambda_1)}; x,\bar{x}\right)
\end{align}
Fixing the conformal scale and appending the rest of the $z$ dependence from (\ref{eq:4graviton1shadow}), we can express the correlator as:
\begin{align}
    G_{34}^{21}(x,\bar{x})=&(1-x)^{3+i\lambda_4}x^{-1+i\lambda_2}(\bar{x})^{3+i\lambda_2}(1-\bar{x})^{-1+i\lambda_4}\nonumber\times\\
    &\times\sum_{n=0}^\infty \frac{(1+i\lambda_2+i\lambda_4)_n(3-i\lambda_1)_n(-1-i\lambda_1)_n}{n!(-1+n+i(\lambda_2-\lambda_1))_n(i\lambda_2-i\lambda_1)_{2n}}\nonumber\times\\
    &\times x^n\ \hy\left({1+i\la_2+i\la_4+n,3-i\la_1+n\atop 2n+i(\la_2-\la_1)};x\right) \nonumber\\
    &\times\bar{x}^n\ \hy\left({1+i\la_2+i\la_4+n,-1-i\la_1+n\atop 2n+i(\la_2-\la_1)};x\right),
\end{align}
where we made use of the following property of the Appell function:
\begin{align}
    F_1\left({a;b_1,b_2\atop c};x,y\right)=\sum_{n=0}^\infty &\fc{(a_n)(b_1)_n(b_2)_n(c-a)_n}{n!(c+n-1)_n(c)_{2n}}x^ny^n\nonumber\\
    &\times\ \hy\left({a+n,b_1+n\atop c+2n};x\right) \hy\left({a+n,b_2+n\atop c+2n};y\right).
\end{align}
So now we have separated the holomorphic and antiholomorphic parts of the correlator, but we need an infinite number of them. The advantage is that we can apply the techniques from the previous section to further decompose each of the two hypergeometric functions into a sum of conformal blocks. After a tedious computation, the block decomposition is the following:
\begin{align}
    G_{34}^{21}(x,\bar{x})=&\sum_{n=0}^\infty \frac{(1+i\lambda_2+i\lambda_4)_n(3-i\lambda_1)_n(-1-i\lambda_1)_n}{n!(-1+n+i(\lambda_2-\lambda_1))_n(i\lambda_2-i\lambda_1)_{2n}}\nonumber\\
    &\times\sum_{k,p=0}^\infty \Tilde{a}_{n,k}\Tilde{b}_{n,p} K_{34}^{21}\left[n+k+1+i\frac{\lambda_2-\lambda_1}{2},n+p+1+i\frac{\lambda_2-\lambda_1}{2}\right](x,\bx),
\end{align}
with the new coefficients being:
\begin{align}
    \Tilde{a}_{n,k}=\frac{\Gamma(2n+i(\lambda_2-\lambda_1))\Gamma(n+k-1-i\lambda_1)}{\Gamma(-1-i\lambda_1+n)\Gamma(1+2(n+k)+i(\lambda_2-\lambda_1))}\bigg[&\frac{\Gamma(2+n+k+i(\lambda_2+\lambda_4))}{\Gamma(n+1+i(\lambda_2+\lambda_4))}\nonumber\\&+(-1)^k\frac{\Gamma(n+k-i(\lambda_1+\lambda_4))}{\Gamma(n-1-i(\lambda_1+\lambda_4))}\bigg] \, ,\nonumber\nonumber\\
    \Tilde{b}_{n,k}=\frac{\Gamma(n+k+3-i\lambda_1)\Gamma(2n+i(\lambda_2-\lambda_1))}{\Gamma(1+2(n+k)+i(\lambda_2-\lambda_1))\Gamma(3+n-i\lambda_1)}\bigg[&\frac{\Gamma(2+n+k+i(\lambda_2+\lambda_4))}{\Gamma(n+1+i(\lambda_2+\lambda_4))}\nonumber\\&+(-1)^k\frac{\Gamma(n+k-i(\lambda_1+\lambda_4))}{\Gamma(n-1-i(\lambda_1+\lambda_4))}\bigg].
\end{align}
We observe that the spectrum contains operators with $(h,\bh)=(n+k+1+\fc{i(\la_2-\la_1)}{2},n+p+1+\fc{i(\la_2-\la_1)}{2})$. They have dimension $\Delta=2n+M+2+i(\la_2-\la_1)$ where $M\geq 0$ is an integer and integer spin $J=-M,-M+2,...,M-2,M$. Note that the exchanged operators depend on $\lambda_2-\lambda_1$ simply because after the shadow transform, $\tilde{\Delta}_1=1-i\lambda_1$ while $\Delta_2=1+i\lambda_2$.

The general conformal block decomposition in the other two channels is more complicated. One might utilize the method developed in \cite{Fan:2023lky} to perform analytic continuation of the shadow correlator. We leave it for future work.


\section{Single-valued shadow correlators} \label{sec3}

\subsection{Single-valued completion in the soft shadow limit}
The shadow correlator in the ``soft'' shadow limit $\lambda_1=i$  (\ref{eq:GS1I1bar}) behaves as $|x|^{2i\lambda_2+6}x^{-4}$ near $x=0$, which has no monodromy. To examine its monodromy near $x= 1$, we analytically continue the hypergeometric function from the holomorphic part to $x\approx 1$:
\begin{equation}
\begin{split}
    S_1(x)=&(1-x)^{3+i\lambda_4}x^{i\lambda_2-1}B(-i\lambda_3,-4-i\lambda_4) \, \hy\left({4,-i\lambda_3\atop 5+i\lambda_4};1-x\right)+\\
    &+\frac{1}{x(1-x)}B(-i\lambda_4,4+i\lambda_4)\, \hy\left({1+i\lambda_3,-3\atop -3-i\lambda_4};1-x\right) \, . \label{eq:S11-x}
\end{split}
\end{equation}
The hypergeometric function in the second term is a rational function due to the presence of a negative parameter $b=-3$. It can also be written as a hypergeometric function with the argument $x$,
\begin{equation}
\begin{split}
    &\frac{1}{x(1-x)}B(-i\lambda_4,4+i\lambda_4)\, \hy\left({1+i\lambda_3,-3\atop -3-i\lambda_4};1-x\right) \\
    =&\frac{1}{x(1-x)}B(-i\lambda_4,4+i\lambda_4) \frac{(4+i\lambda_3+i\lambda_4)(3+i\lambda_3+i\lambda_4)(2+i\lambda_3+i\lambda_4)}{(3+i\lambda_4)(2+i\lambda_4)(1+i\lambda_4)} \times \\
    &\qquad \times\, \hy\left({-3,1+i\lambda_3\atop 1-i\lambda_2}; x\right) \, . \label{eq:S11-xsec} 
\end{split}
\end{equation}
Therefore, the shadow correlator (\ref{eq:GS1I1bar}) contains  a term that behaves as $(1-\bar{x})^{-1+i\lambda_4}/(1-x)$ near $x=1$ and has a nontrivial monodromy. Note that the contribution of the first term in (\ref{eq:S11-x}) to $S_1(x)\bar{I}_1(x)$ is single-valued near $x=1$. Therefore, to construct a single-valued shadow correlator, we need to supplement (\ref{eq:GS1I1bar}) with additional contributions. 

Recall that the antiholomphic part (\ref{eq:antisingleblock}) is described by a single block with $\bar{h}=\frac{3}{2}+\frac{i\lambda_2}{2}$. Following the constructions shown in \cite{Fan:2021pbp}, we consider the shadow block with $\bar{h}'=1-\bar{h}=-\frac{1}{2}-\fc{i\lambda_2}{2}$,
\begin{equation}
    \bar{I}_2(\bar{x})=\bar{x}^{1-\bar{h}-\bar{h}_3-\bar{h}_4} \,_2F_1\left({ 1-\bar{h}-\bar{h}_{12},\, 1-\bar{h}+\bar{h}_{34}\atop 2-2\bar{h}};\bar{x}\right)=\bar{x}\ \hy\left({1, i\lambda_3\atop -1-i\lambda_2};\bar{x}\right)
\end{equation}
Analytically continuing to $\bar{x}=1$ we get:
\begin{equation}
\bar{I}_2(\bar{x})=\bar{x}\frac{2+i\lambda_2}{(1-i\lambda_4)}\hy\left({1,i\lambda_3\atop 2-i\lambda_4};1-\bar{x}\right)+\bar{x}^{1-i\lambda_2}(1-\bar{x})^{-1+i\lambda_4}\frac{\Gamma(-1-i\lambda_2)\Gamma(1-i\lambda_4)}{\Gamma(i\lambda_3)}.
\end{equation}
Notice that the second term shares the same antiholomorphic factor (\ref{nonholschan}) with the term that we would like to cancel from (\ref{eq:S11-xsec}). It is clear that near $x=1$, the following combination is single-valued,
\begin{equation}
    G^{21}_{34}(x,\bar{x})_{SV}=S_1(x) \bar{I}_1(\bar{x}) + S_2(x)\bar{I}_2(\bar{x}) \, , \label{eq:GshadowSV}
\end{equation}
with 
\begin{equation}
\begin{split}
    S_2(x)=&\frac{B(i\lambda_3,i\lambda_4)}{6}(2+i\lambda_3+i\lambda_4)(3+i\lambda_3+i\lambda_4)(4+i\lambda_3+i\lambda_4)\times\\
    &\times\frac{1}{x(1-x)} \, \hy\left({-3,1+i\lambda_3\atop 1-i\lambda_2}; x\right)\, . \label{eq:S2}
\end{split}
\end{equation}

One can check that this combination is also single-valued near $x=\infty$.

\subsection{Conformal block decomposition} \label{sec:3.3}
We have shown that the single-valued shadow correlator (\ref{eq:GshadowSV}) is defined over the entire complex plane. By construction, it satisfies the crossing symmetry constraints,
\begin{equation}
    G_{32}^{41}(1-x,1-\bx)_{SV}=G_{34}^{21}(x,\bx)_{SV}\ ,\qquad G_{31}^{24}\left(\frac{1}{x},\frac{1}{\bx}\right)_{SV}=x^{2h_3}\bar{x}^{2\bh_3}
G_{34}^{21}(x,\bx)_{SV}\ .
\end{equation}
We would now like to perform the conformal block decomposition of the single-valued shadow correlator (\ref{eq:GshadowSV}).
We start with the $(12\rightleftharpoons 34)_{\mathfrak{2}}$ channel. The $S_1(x)\bar{I}_1(\bar{x})$ has already been computed in a previous section and we constructed $\bar{I}_2(\bar{x})$ as a single conformal block. It remains to decompose $S_2(x)$. Similarly to $S_1(x)$, it does not describe a single block, but can be described as an infinite sum. The final result is:
\begin{multline}
    G_{34}^{21}(x,\bar{x})_{SV}=\sum_{k=1}^\infty\left( a_k K_{34}^{21}\left[k+\frac{1}{2}+\frac{i\lambda_2}{2},\frac{3}{2}+\frac{i\lambda_2}{2}\right](x,\bx)+\right.\\
     \left. + b_k K_{34}^{21}\left[k+\frac{1}{2}-\frac{i\lambda_2}{2},-\frac{1}{2}-\frac{i\lambda_2}{2}\right](x,\bx)\right) \, ,
    \label{eq:SVconfblockschan}
\end{multline}
where $a_k$ is given in (\ref{ak_coeff}) and $b_k$ is a new coefficient given below:
\be
    b_k=\frac{\Gamma(1+ i\lambda_4+k)\Gamma(3-i\lambda_2+k)\Gamma(i\lambda_3)}{\Gamma(-i\lambda_2+2k)\Gamma(-1-i\lambda_2)\Gamma(4)i\lambda_4}+\frac{(-1)^{k-1}\Gamma(1+ i\lambda_3+k)\Gamma(3-i\lambda_2+k)\Gamma(i\lambda_4)}{\Gamma(-i\lambda_2+2k)\Gamma(-1-i\lambda_2)\Gamma(4)i\lambda_3}.
\ee
\par One of the issues identified in the initial shadow correlator was that, when continued to $x\approx 1$, and decomposed into its conformal blocks, the spin of the exchanged particles was continuous, depending on $i\lambda_2$. This issue is resolved for the single-valued correlator as follows. Analytically continuing (\ref{eq:GshadowSV}) to $x\approx 1$ the correlator becomes:
\begin{equation}
\begin{split}
    &G_{32}^{41}(1-x,1-\bar{x})_{SV}\\
    =&(1-x)^{3+i\lambda_4}x^{i\lambda_2-1}B(-i\lambda_3,-4-i\lambda_4)\ \hy\left({4,-i\lambda_3\atop 5+i\lambda_4};1-x\right)(1-\bar{x})^{-1+i\lambda_4}\bar{x}^{3+i\lambda_2}+\\
    +&\bar{x}\ \hy\left({1,i\lambda_3\atop 2-i\lambda_4};1-\bar{x}\right)\times\\
    &\times\frac{x^{i\lambda_2-1}}{1-x}\frac{\Gamma(i\lambda_3)\Gamma(4+i\lambda_4)(2+i\lambda_2)}{\Gamma(-1-i\lambda_2)\Gamma(4)(i\lambda_4)(1-i\lambda_4)}\ \hy\left({-i\lambda_4,-3+i\lambda_2\atop -3-i\lambda_4};1-x\right).
    \label{eq:x=1corr}
\end{split}
\end{equation}
In the $(14\rightleftharpoons 32)_{\mathfrak{2}}$ channel, a conformal block has the decomposition:
\begin{equation}
    \begin{split}
    K_{32}^{41}[h,\bar{h}](1-x,1-\bar{x})=&(1-x)^{h-h_3-h_2}\ \hy\left({h-h_{14},h+h_{32}\atop 2h};1-x\right)\times\\ &(1-\bar{x})^{\bar{h}-\bar{h}_3-\bar{h}_2}\ \hy\left({\bar{h}-\bar{h}_{14},\bar{h}+\bar{h}_{32}\atop 2\bar{h}};1-\bar{x}\right),
\end{split}
\end{equation}
with:
\begin{align}\nonumber&
h_{14}= \textstyle\frac{1}{2}- \frac{i\lambda_4}{2}\ , & \bh_{12}=\textstyle\frac{1}{2}-\frac{i\lambda_4}{2}\ ,\\[1mm] &
h_{32}=\textstyle2+\frac{i\lambda_3}{2}-\textstyle\frac{i\lambda_2}{2}
\ , & \bh_{32}=-2+\textstyle\frac{i\lambda_3}{2}-\textstyle\frac{i\lambda_4}{2}\ ,\\[1mm] &
h_3+h_2=1+\textstyle\frac{i\lambda_3}{2}+\textstyle\frac{i\lambda_2}{2}\ , & ~~~~~~\bh_3+\bh_2=1+\textstyle\frac{i\lambda_3}{2}+\textstyle\frac{i\lambda_4}{2}.\nonumber
\end{align}
To decompose the two correlator contributions, we apply the identities in (\ref{hy1}-\ref{hy3}) to (\ref{eq:x=1corr}). The final decomposed correlator is:
\be
\begin{split}
    G_{32}^{41}(1-x,1-\bar{x})_{SV}=&\sum_{k=1}^\infty c_k K_{32}^{41}\left[\frac{5}{2}+k+\frac{i\lambda_4}{2},-\frac{1}{2}+\frac{i\lambda_4}{2}\right](1-x,1-\bar{x})+\\
    &+\sum_{k=1}^\infty d_k K_{32}^{41}\left[-\frac{3}{2}+k-\frac{i\lambda_4}{2},\frac{1}{2}-\frac{i\lambda_4}{2}\right](1-x,1-\bar{x})+\\
    &+\sum_{k=1}^\infty \frac{3+i\lambda_2}{2-i\lambda_4}d_k K_{32}^{41}\left[-\frac{3}{2}+k-\frac{i\lambda_4}{2},\frac{3}{2}-\frac{i\lambda_4}{2}\right](1-x,1-\bar{x}).
\end{split}
\ee
The two coefficients that appear in the sum are:
\be
\begin{split}
    c_k=&\fc{\sin(\pi i(\la_3+\la_4))\Gamma(3+k)}{6\sin(\pi i\la_4)\Gamma(4+2k+i\la_4)}\left[\Gamma(k-i\la_3)\Gamma(5+i\la_3+i\la_4)+(-1)^{k+1}\Gamma(-i\la_3)\Gamma(5+k+i\la_3+i\la_4)\right]\\
    d_k=&\fc{\Gamma(-3-i\la_4)\Gamma(k-i\la_4-1)\Gamma(4+i\la_4)\Gamma(i\la_3)}{6\Gamma(2-i\la_4)\Gamma(-4+2k-i\la_4)\Gamma(1+i\la_3)\Gamma(-1+i\la_3+i\la_4)}\Big[(-1)^{k+1}\Gamma(1+k+i\la_3)+\\
    &+\fc{\Gamma(-4+k-i\la_3-i\la_4)\Gamma(1+i\la_3)}{\Gamma(-4-i\la_3-i\la_4)}\Big].
\end{split}
\ee
Interestingly, in the $(14\rightleftharpoons 32)_{\mathfrak{2}}$ channel, we find three sets of conformal blocks. The first comes with $\Delta=J-1+i\lambda_4$, with $J\geq 4$ an integer spin. The second set has dimensions $\Delta=J+1-i\lambda_4$ with $J\geq -1$ being an integer spin. Finally, the third block described fields with $\Delta=J+3-i\lambda_4$ and integer spin $J\geq -2$. The $J=-2$ contribution of this block describes a negative helicity graviton appearing in the product $G_{\Delta_2}^{-,-\epsilon}(1,1)G_{\Delta_3}^{+,\epsilon}(x,\bar{x})$ as we will show later in this section. We also observe that, as claimed earlier, there are no exotic states with non-integer spins appearing in the single-valued correlator.  
\par For completeness, we can perform the same analysis for the $(13\rightleftharpoons 42)_{\mathfrak{2}}$ block. Here, a conformal block of a primary field is given by:
\be
\begin{split}
    K_{31}^{24}[h,\bar{h}]\left(\frac{1}{x},\frac{1}{\bar{x}}\right) =& \left(\frac{1}{x}\right)^{h-h_3-h_2}\ \hy\left({h-h_{42},h+h_{31}\atop 2h};\frac{1}{x}\right)\times\\
    &\times\left(\frac{1}{x}\right)^{\bh-\bh_3-\bh_2}\ \hy\left({\bh-\bh_{42},\bh+\bh\atop 2\bh};\frac{1}{\bar{x}}\right)
\end{split}
\ee
where this time:
\begin{align}\nonumber&
h_{42}= \textstyle2+ \frac{i\lambda_4}{2}-\frac{i\lambda_2}{2}\ , & \bh_{42}=\textstyle-2-\frac{i\lambda_4}{2}-\frac{i\lambda_2}{2}\ ,\\[1mm] &
h_{31}=-\textstyle\frac{1}{2}+\frac{i\lambda_3}{2}
\ , & \bh_{31}=-\textstyle\frac{1}{2}+\textstyle\frac{i\lambda_3}{2}\ ,\\[1mm] &
h_3+h_1'=\textstyle\frac{7}{2}+\textstyle\frac{i\lambda_3}{2}\ , & ~~~~~~\bh_3+\bh_1'=-\textstyle\frac{1}{2}+\textstyle\frac{i\lambda_3}{2}.\nonumber
\end{align}
The analytically continued correlator is:

\begin{multline}
    G^{24}_{31}\left(\frac{1}{x},\frac{1}{\bar{x}}\right)_{SV}=\left(1-\frac{1}{x}\right)^{3+i\lambda_4}B(-i\lambda_4,-i\lambda_3-4)\ \hy\left({4,4-i\lambda_2\atop 5+i\lambda_3};\fc{1}{x}\right)\left(1-\frac{1}{\bar{x}}\right)^{-1+i\lambda_4}\\
    +\frac{x^{5+i\lambda_3}}{1-x}\frac{\Gamma(i\lambda_4)\Gamma(4+i\lambda_3)}{(i\lambda_3)(1-i\lambda_3)\Gamma(4)\Gamma(-2-i\lambda_2)}\hy\left({-3,-3+i\lambda_2\atop -3-i\lambda_3};\fc{1}{x}\right)\\
    \times\bar{x}^{-1+i\lambda_3}\ \hy\left({1,3+i\lambda_2\atop 2-i\lambda_3};\fc{1}{\bx}\right).
\end{multline}

The block decomposition then reads:
\be
\begin{split}
G^{24}_{31}\left(\frac{1}{x},\frac{1}{\bar{x}}\right)_{SV}=&\sum_{k=1}^\infty e_k K_{31}^{24}\left[k+\frac{5}{2}+\frac{i\lambda_3}{2},-\fc{1}{2}+\fc{i\lambda_3}{2}\right]\left(\fc{1}{x},\fc{1}{\bx}\right)+\\
+&\sum_{k=1}^\infty f_k K_{31}^{24}\left[k-\fc{3}{2}-\fc{i\la_3}{2},\fc{1}{2}-\fc{i\la_3}{2}\right]\left(\fc{1}{x},\fc{1}{\bx}\right)+\\
+&\sum_{k=1}^\infty \fc{3+i\la_2}{2-i\la_3} f_k K_{31}^{42}\left[k-\fc{3}{2}-\fc{i\la_3}{2},\fc{3}{2}-\fc{i\la_3}{2} \right]\left(\fc{1}{x},\fc{1}{\bx}\right),
\end{split}
\ee
with 
\be
\begin{split}
    e_k=&\fc{\sin(\pi(i\la_3+i\la_4))\Gamma(3+k)}{6\sin(\pi i\la_3)\Gamma(4+2k+i\la_3)}\left[\Gamma(-i\la_4)\Gamma(5+i\la_3+i\la_4+k)+(-1)^{k+1}\Gamma(k-i\la_4)\Gamma(5+i\la_3+i\la_4)\right]\\
    f_k=&\fc{\Gamma(-3-i\la_3)\Gamma(k-i\la_3-1)\Gamma(4+i\la_3)\Gamma(i\la_4)}{6\Gamma(2-i\la_3)\Gamma(-4+2k-i\la_3)\Gamma(1+i\la_4)\Gamma(-1+i\la_3+i\la_4)}\Big[\Gamma(1+k+i\la_4)+\\
    &+(-1)^{k+1}\fc{\Gamma(-4+k-i\la_3-i\la_4)\Gamma(1+i\la_4)}{\Gamma(-4-i\la_3-i\la_4)}\Big].
\end{split}
\ee
As in the $(14\rightleftharpoons32)_{\mathfrak{2}}$ channel, we see the appearance of 3 classes of blocks. They all describe fields with integer spin. The first block is described by $\Delta=J-1+i\la_3$ and $J\geq 4$, the second by $\Delta=J+1-i\la_3$ with $J\geq -1$ and the last one by $\Delta=J+3-i\la_3$ with $J\geq2$. We observe that the spectrum is identical to that in the  $(14\rightleftharpoons32)_{\mathfrak{2}}$ channel, after exchanging the indices 3 and 4. This is a consequence of the crossing symmetry. 

We would like to compare the leading OPEs of the single-valued shadow correlator (\ref{eq:GshadowSV}) to the known celestial OPEs of gravitons derived in \cite{Pate:2019lpp}. In the $(12\rightleftharpoons34)_{\mathfrak{2}}$ channel, $x\approx0$ corresponds to the limit $z_3\rightarrow z_4$. The leading OPE of two positive helicity gravitons from the single-valued shadow correlator is 
\begin{equation}
    G^{+,\epsilon}_{\Delta_3}(z_3,\bar{z}_3)G^{+,\epsilon}_{\Delta_4}(z_4,\bar{z}_4) \sim \frac{\bar{z}_{34}}{z_{34}}B(\Delta_3-1,\Delta_4-1)G^{+,\epsilon}_{\Delta_3+\Delta_4}(z_4,\bar{z}_4) \, , \label{eq:G+G+}
\end{equation}
where $\epsilon=1$ indicating the particle is outgoing in our case. The leading OPE (\ref{eq:G+G+}) agrees with the well-known celestial OPE \cite{Pate:2019lpp}.

In the $(14\rightleftharpoons 32)_{\mathfrak{2}}$ channel, $x\approx1$ corresponds to the limit $z_3\rightarrow z_2$. The OPE of two gravitons with opposite helicities from the single-valued shadow correlator is
\begin{equation}
\begin{split}
    G_{\Delta_3}^{+,\epsilon}(z_3,\bar{z}_3) G_{\Delta_2}^{-,-\epsilon}(z_2,\bar{z}_2) &\sim \frac{1}{z_{32}}B(\Delta_3-1,-\Delta_3-\Delta_2)O^{-\epsilon}_{\Delta_3+\Delta_2-1,J=-1}(z_2,\bar{z}_2) +\\
    &+\frac{\bar{z}_{32}}{z_{32}}B(\Delta_3-1,-1-\Delta_3-\Delta_2)G_{\Delta_3+\Delta_2}^{-,-\epsilon}(z_2,\bar{z}_2) \, . \label{eq:G+G-}
\end{split}
\end{equation}
While the second line in (\ref{eq:G+G-}) agrees with the one in \cite{Pate:2019lpp}, we found a new operator $O^{-\epsilon}_{\Delta_3+\Delta_2-1,J=-1}(z_2,\bar{z}_2)$ appearing in the OPE. On a technical level, its presence is a consequence of the shadow correlator single-valuedness (\ref{eq:GshadowSV}). However, the appearance of the operator in the OPE also suggests that the translation symmetry corresponding to soft gravitons is broken in the single-valued correlator. This can happen if, for instance, the bulk space-time origin of the correlator is gravity placed on a non-trivial background. We suspect this operator might correspond to photons coupled to gravity based on the spin of the operator that appears on the R.H.S of  (\ref{eq:G+G-}). We leave making such a connection precise for future work.

Finally, in the $(13\rightleftharpoons 24)_{\mathfrak{2}}$ channel, the OPE can be obtained from (\ref{eq:G+G-}) by replacing $3$ by $4$ as the single-valued shadow correlator is manifestly symmetric under $3\leftrightarrow4$.

\subsection{Integral representation of the single-valued shadow correlator}
The single-valued correlator (\ref{eq:GshadowSV}) resembles the four-point correlator in the minimal models computed by the Coulumb gas formulation \cite{Dotsenko:1984ad,Dotsenko:1984nm}, where the correlator can be expressed in terms of a complex integral over a single-valued function. Therefore, we expect (\ref{eq:GshadowSV}) to have such an integral representation. The integrals considered in \cite{Dotsenko:1984ad,Dotsenko:1984nm} in the context of four-point correlators in minimal models have the form:
\be
\mathcal{I}(x,\bar{x})=\int d^2 w\ w^{\hat{
a}+a}\bar{w}^{\bar{a}+a}(w-1)^{\hat{b}+b}(\bar{w}-1)^{\bar{b}+b}(w-x)^{\hat{c}+c}(\bar{w}-x)^{\bar{c}+c},
\ee
where $a,b,c,\bar{a},\bar{b},\bar{c}\in\mathbb{Z}$ and $\hat{a},\hat{b},\hat{c}\notin \mathbb{Z}$. The integral can be evaluated by separating the holomorphic and anti-holomorphic contributions, resulting in:
\begin{align}
{\cal I}(x,\bar x)&=\frac{s(\hat b)s(\hat a+\hat b+\hat c)}{s(\hat a+\hat c)}\
{\cal I}_1(\hat a+a,\hat b+b,\hat c+c;x)\ {\cal I}_1(\hat a+\bar a,\hat b+\bar b,\hat c+\bar c;\bar x)\nonumber\\
&+\ (-1)^{b+\bar b+c+\bar c}\ \frac{s(\hat a)s(\hat c)}{s(\hat a+\hat c)}\  {\cal I}_2(\hat a+a,\hat b+b,\hat c+c;x)\ {\cal I}_2(\hat a+\bar a,\hat b+\bar b,\hat c+\bar c;\bar x)\ ,\label{i12}
\end{align}
with $s(x)\equiv \sin(\pi x)$ and
\be
\begin{split}
    &\mathcal{I}_1(a,b,c;x)=B(-a-b-c-1,b+1)\ \hy\left({-c,-a-b-c-1\atop -a-c};x\right)\\
    &\mathcal{I}_2(a,b,c;x)=x^{a+c+1} B(a+1,c+1)\ \hy\left({-b,1+a\atop a+c+2};x\right).
\end{split}
\ee

By inspecting the form of (\ref{eq:GshadowSV}), we can see that it has the following integral representation:
\begin{equation}
    G_{34}^{21}(x,\bar{x})_{SV}=\frac{1}{\pi}(2+i\lambda_2)\ B(i\lambda_3,i\lambda_4)\ \mathcal{I}(x,\bar{x})\ \frac{\bar{x}}{x(1-x)},
    \label{eq:IntrepSV}
\end{equation}
where the coefficients that define $\mathcal{I}(x,\bar{x})$ are $\hat{a}=-i\lambda_4,\ \hat{b}=-i\lambda_2,\ \hat{c}=-i\lambda_3;\ a=-1,\ b=3,\ c=-1;\ \bar{a}=0,\ \bar{b}=-3,\ \bar{c}=0$.  The holomorphic factors of the integrand
have exponents that differ by integers from the antiholomorphic ones, resembling the structure found in \cite{Fan:2021pbp} for the single-valued shadow correlator of four gluons.

\section{Single-valued celestial graviton amplitude from inverted shadow} \label{sec4}
In the previous section we wrote the single-valued correlator (\ref{eq:GshadowSV}) as a complex integral. In \cite{Fan:2021pbp} a similar integral representation was found for the single-valued completion of the four-gluon correlator with one shadow field. There it was interpreted as a shadow transform of a different single-valued celestial amplitude. In the present section we will show that our integral representation can also be recast as a shadow transform after a suitable change of variables. We recall that we initially chose to perform a shadow transform on the celestial graviton correlator in order to eliminate the distributional support on the celestial sphere. By reverting the shadow transform of the new amplitude (\ref{eq:GshadowSV}) we will land on a single-valued correlator that is more natural from a CFT perspective.
\par In order to find a suitable change of variables inside the integral representation (\ref{eq:IntrepSV}), we will need to reinstall the full coordinate depedence of the four-graviton correlator:
\begin{align}
    \langle \tilde{\phi}_{\tilde{\Delta}_1,+}(z_{1'},\bar{z}_{1'})&\phi_{\Delta_2,-}(z_2,\bar{z}_2)\phi_{\Delta_3,+}(z_3,\bar{z}_3)\phi_{\Delta_4,+}(z_4,\bar{z}_4)\rangle = \nonumber\\
    &(z_{1'2})^{1-i\lambda_2}(z_{1'3})^{-3-i\lambda_3}(z_{1'4})^{-3-i\lambda_4}(\bar{z}_{1'2})^{-3-i\lambda_2}(\bar{z}_{1'3})^{1-i\lambda_3}(\bar{z}_{1'4})^{1-i\lambda_4}G^{21}_{34}(x,\bar{x}),
    \label{eq:SVcorrwithweights}
\end{align}
where we also returned to the original notation of the shadowed gluon coordinates $z_1'$. Note that $\tilde{\Delta}_1=2-\Delta_1=1$ in this case. In terms of this coordinate and the position before the shadow transform, we can express the invariant ratios as:
\begin{equation}
    x=\frac{z_{1'2}z_{34}}{z_{1'3}z_{24}},\ z=\frac{z_{12}z_{34}}{z_{13}z_{24}}.
\end{equation}
\par The integral in (\ref{eq:IntrepSV}) is expressed in terms of the variable $w$. We would like to find a suitable relation between $w$ and $z_1$ such that after the change of variables, the integral measure will become:
\begin{equation}
    \int \fc{\mathrm{d}^2 z_1}{z_{11'}^4}
\end{equation}
and the dependence of $G_{34}^{21}(x,\bar{x})$ on the differences $z_{1'2},z_{1'3},z_{1'4}$ and their complex conjugates cancels the prefactor of (\ref{eq:SVcorrwithweights}). These conditions are enough to single out the following relationship between $w$ and $z_{1'}$:
\begin{equation}
    w=\fc{z_{14}z_{21'}}{z_{11'}z_{24}}.
\end{equation}
This further enables us to express the other quantities appearing in the integral representation and the Jacobian as:
\begin{equation}
    w-1=-\fc{z_{12}z_{1'4}}{z_{11'}z_{24}},\quad w-x=\fc{z_{21'}z_{13}z_{1'4}}{z_{24}z_{11'}z_{1'3}},\quad \fc{\mathrm{d}w}{\mathrm{d}z_1}=\fc{z_{1'2}z_{1'4}}{z_{11'}^2z_{24}}.
\end{equation}
The integral in the correlator becomes:
\begin{align}
    &\frac{\bar{x}}{x(1-x)}\int \mathrm{d}^2 w^{-1-i\lambda_4}\bar{w}^{-i\lambda_4}(w-1)^{3-i\lambda_2}(\bar{w}-1)^{-3-i\lambda_2}(w-x)^{-1-i\lambda_3}(\bar{w}-\bar{x})^{-i\lambda_3}=\nonumber\\
    &=\frac{\bar{z}_{34}}{z_{24}z_{34}z_{23}}z_{1'2}^{-1+i\lambda_2}z_{1'3}^{3+i\lambda_3}z_{1'4}^{3+i\lambda_4}(\bar{z}_{1'2})^{3+i\lambda_2}(\bar{z}_{1'3})^{-1+i\lambda_3}(\bar{z}_{1'4})^{-1+i\lambda_4}\nonumber\\
    &\int  \frac{\mathrm{d}^2 z_1}{z_{11'}^4} (z_{12})^{3-i\lambda_2}(z_{13})^{-1-i\lambda_3}(z_{14})^{-1-i\lambda_4}(\bar{z}_{12})^{-3-i\lambda_2}(\bar{z}_{13})^{-i\lambda_3}(\bar{z}_{14})^{-i\lambda_4}.
\end{align}
With this reparametrization, the correlator (\ref{eq:SVcorrwithweights}) can be rewritten as:
\begin{align}
     \langle \tilde{\phi}_{\tilde{\Delta}_1,+}(z_{1'},\bar{z}_{1'})&\phi_{\Delta_2,-}(z_2,\bar{z}_2)\phi_{\Delta_3,+}(z_3,\bar{z}_3)\phi_{\Delta_4,+}(z_4,\bar{z}_4)\rangle \nonumber\\
     &\int \fc{\mathrm{d}^2 z_1}{z_{11'}^4}\langle \phi_{\Delta_1,-}(z_1,\bar{z}_1)\phi_{\Delta_2,-}(z_2,\bar{z}_2)\phi_{\Delta_3,+}(z_3,\bar{z}_3)\phi_{\Delta_4,+}(z_4,\bar{z}_4)\rangle_{SV},
\end{align}
where the single-valued celestial amplitude is:
\begin{align}
    \langle \phi_{\Delta_1,-}(z_1,\bar{z}_1)&\phi_{\Delta_2,-}(z_2,\bar{z}_2)\phi_{\Delta_3,+}(z_3,\bar{z}_3)\phi_{\Delta_4,+}(z_4,\bar{z}_4)\rangle_{SV}=\frac{(2+i\lambda_2)B(i\lambda_3,i\lambda_4)}{\pi}\times\nonumber\\
    &\times\frac{\bar{z}_{34}}{z_{24}z_{34}z_{23}}(z_{12})^{3-i\lambda_2}(z_{13})^{-1-i\lambda_3}(z_{14})^{-1-i\lambda_4}(\bar{z}_{12})^{-3-i\lambda_2}(\bar{z}_{13})^{-i\lambda_3}(\bar{z}_{14})^{-i\lambda_4}.
\end{align}
We use conformal invariance to simplify the expression to:
\begin{align}
    \mathcal{G}_{34}^{21}(z,\bar{z})_{GR}=&\lim_{z_1\rightarrow\infty}\langle \phi_{\Delta_1,-}(z_1,\bar{z}_1)\phi_{\Delta_2,-}(z_2,\bar{z}_2)\phi_{\Delta_3,+}(z_3,\bar{z}_3)\phi_{\Delta_4,+}(z_4,\bar{z}_4)\rangle_{SV}=\nonumber\\
    &=\frac{(2+i\lambda_2)B(i\lambda_3,i\lambda_4)}{\pi} \frac{\bar{z}}{z(1-z)}. \label{eq:invertGR}
\end{align}
\par Let us compare the conformal block decomposition of the single-valued celestial correlator just obtained and the initial one studied earlier in $\ref{sec:confblockdecomp}$. For the $(12\rightleftharpoons34)_{\mathfrak{2}}$ channel, we have a single anti-holomorphic block and an infinite sum of holomorphic ones:
\begin{equation}
    \mathcal{G}_{34}^{21}(z,\bar{z})=\sum_{k=1}^\infty g_k K_{34}^{21}\left[k+\fc{1}{2}-\fc{i\la_2}{2},-\fc{1}{2}-\fc{i\la_2}{2}\right](z,\bar{z})
\end{equation}
where:
\begin{equation}
    g_k=\frac{(2+i\lambda_2)B(i\lambda_3,i\lambda_4)}{\pi}\frac{\Gamma(1-i\lambda_2+k)}{\Gamma(2-i\lambda_2+2k)}\left( \frac{(-1)^k\Gamma(2+k+i\lambda_3)}{\Gamma(1+i\lambda_3)}+\frac{\Gamma(2+k+i\lambda_4)}{\Gamma(1+i\lambda_4)}\right).
\end{equation}
The spectrum consists of fields with integer spin $J\geq 2$ and dimension $\Delta=J-1-i\la_2$. The leading contribution is given by a field with spin $+2$ and dimension $\Delta=1-i\la_2=2+i\la_3+i\la_4$. In contrast to the (\ref{eq:SVconfblockschan}) block decomposition, we observe that the current correlator has a single set of blocks. 

\par For the $(14\rightleftharpoons 32)_{\mathfrak{2}}$ channel, the block decomposition reads:
\begin{align}
    \mathcal{G}_{14}^{32}(1-z,1-\bar{z})=&\sum_{k=1}^\infty h_k K_{14}^{32}\left[-\fc{3}{2}+k-\fc{i\la_4}{2},\fc{1}{2}-\fc{i\la_4}{2}\right](1-z,1-\bar{z})+\nonumber \\
    +&\sum_{k=1}^\infty \fc{3+i\la_2}{1-i\la_4}h_k K_{14}^{32}\left[-\fc{3}{2}+k-\fc{i\la_4}{2},\fc{3}{2}-\fc{i\la_4}{2}\right](1-z,1-\bar{z}),
\end{align}
with 
\begin{equation}
    h_{k}=\fc{(2+i\la_2)\Gamma(-4-i\la_4+k)B(i\la_3,i\la_4)}{\pi\Gamma(-4-i\la_4+2k)}\left(\fc{\Gamma(-4+k-i\la_3-i\la_4)}{\Gamma(-4-i\la_3-i\la_4)}+(-1)^{k+1} \fc{\Gamma(1+k+i\la_3)}{\Gamma(1+i\la_3)}\right).
\end{equation}
For the $(13\rightleftharpoons 24)_{\mathfrak{2}}$ channel, the block decomposition is:
\begin{align}
    \mathcal{G}_{13}^{24}\left(\fc{1}{z},\fc{1}{\bar{z}}\right)=&\sum_{k=1}^\infty j_k K_{13}^{24}\left[-\fc{3}{2}+k-\fc{i\la_3}{2},\fc{1}{2}-\fc{i\la_3}{2}\right]\left(\fc{1}{z},\fc{1}{\bar{z}}\right)+\nonumber\\
    +&\sum_{k=1}^\infty \fc{3+i\la_2}{1-i\la_3} j_k K_{13}^{24}\left[-\fc{3}{2}+k-\fc{i\la_3}{2},\fc{3}{2}-\fc{i\la_3}{2}\right]\left(\fc{1}{z},\fc{1}{\bar{z}}\right),
\end{align}
with the coefficient:
\begin{equation}
    j_k=-\fc{(2+i\la_2)B(i\la_3,i\la_4)}{\pi}\left(\fc{\Gamma(1+k+i\la_3)}{\Gamma(1+i\la_3)}+(-1)^{k+1}\fc{\Gamma(-4+k-i\la_3-i\la_4)}{\Gamma(-4-i\la_3-i\la_4)}\right).
\end{equation}
\subsection{Double copy structure}
One way of seeing this simple expression (\ref{eq:invertGR}) is through the results for the gluon correlator from \cite{Fan:2021pbp}. There,  the single-valued gluon correlator is given by a sum of two color ordered contributions:
\begin{equation}
\begin{split}
    G_{34}^{21}(z,\bar{z})_{YM}&= \frac{(1+i\lambda_2)B(i\lambda_3,i\lambda_4)}{\pi z(1-z)} (f^{a_1a_2b}f^{a_3a_4b}-zf^{a_1a_3b}f^{a_2a_4b}) \\
    &=f^{a_1a_2b}f^{a_3a_4b}\tilde{A}_4(1234)+ f^{a_1a_3b}f^{a_2a_4b} \tilde{A}_4(1324) \, ,\label{eq:invertYM}
\end{split}
\end{equation}
where $\tilde{A}_4$s are the color-ordered single-valued celestial four-point gluon amplitudes \cite{Fan:2021pbp}.

The perturbative calculations of amplitudes in gravity and Yang-Mills theory exhibit a remarkable double-copy structure \cite{Kawai:1985xq,Bern:2008qj}, known as Kawai-Lewellen-Tye (KLT) relations \cite{Kawai:1985xq} and Bern-Carrasco-Johansson (BCJ) double copy \cite{Bern:2008qj}. The double-copy structure of celestial amplitudes were studied in \cite{Casali:2020vuy,Casali:2020uvr}. See also \cite{Stieberger:2018onx} for a relation between celestial Einstein-Yang-Mills amplitude with a single graviton and celestial gluon amplitudes. Here, we shall see that our single-valued celestial amplitudes also exhibit a double copy structure.

Recall that the flat space four-point double copy relation between graviton and gluon amplitudes at the tree-level expressed in momentum basis is given by
\begin{equation}
    M_4[1234]=s_{34}A_4[1234]A_4[1243] \, ,
\end{equation}
where $M_4$ is a four-graviton tree amplitude while $A_4$s are color-ordered tree amplitudes in Yang-Mills and $s_{34}=(p_3+p_4)^2$ is the Mandelstam variable $s$.

In (\ref{eq:invertGR}), we observe that $\fc{\bar{z}}{z(1-z)}$ can be expressed as the product of the Parke-Taylor denominators, appearing both in the flat space amplitudes and the celestial correlators, and the two-dimensional kinematic factor $s_{34}=z\bar{z}$. Thus, the single-valued celestial graviton amplitude (\ref{eq:invertGR}) can be understood, up to numerical constants, as the double copy of the four-point single-valued celestial gluon amplitude (\ref{eq:invertYM}),
\begin{equation}
    \mathcal{G}_{34}^{21}(z,\bar{z})_{GR}\sim z \bar{z} \tilde{A}_4(1234) \tilde{A}_4(1243) \, .
\end{equation}
The double copy structure is reminiscent from the Sugawara construction in \cite{Fan:2020xjj}. It would be interesting to make a more precise connection to \cite{Fan:2020xjj} by considering Einstein-Yang-Mills
(EYM) amplitudes.

\subsection{Celestial OPEs}
Similar to what we did in section \ref{sec:3.3}, we would like to compare the leading OPEs of the single-valued correlator (\ref{eq:invertGR}) to the known celestial OPEs of gravitons derived in \cite{Pate:2019lpp}. In the $(12\rightleftharpoons34)_{\mathfrak{2}}$ channel, $x\approx0$ corresponds to the limit $z_3\rightarrow z_4$. The leading OPE of two positive helicity gravitons from the single-valued correlator is 
\begin{equation}
    G^{+,\epsilon}_{\Delta_3}(z_3,\bar{z}_3)G^{+,\epsilon}_{\Delta_4}(z_4,\bar{z}_4) \sim \frac{\bar{z}_{34}}{z_{34}}B(\Delta_3-1,\Delta_4-1)G^{+,\epsilon}_{\Delta_3+\Delta_4}(z_4,\bar{z}_4) \, , \label{eq:G+G+SV}
\end{equation}
where $\epsilon=1$ indicating the particle is outgoing in our case. The leading OPE (\ref{eq:G+G+SV}) agrees with the one in \cite{Pate:2019lpp}.

In the $(14\rightleftharpoons 32)_{\mathfrak{2}}$ channel, $x\approx1$ corresponds to the limit $z_3\rightarrow z_2$. The OPE of two gravitons with opposite helicities from the single-valued correlator is
\begin{equation}
\begin{split}
    G_{\Delta_3}^{+,\epsilon}(z_3,\bar{z}_3) G_{\Delta_2}^{-,-\epsilon}(z_2,\bar{z}_2) &\sim \frac{1}{z_{32}}B(\Delta_3-1,-\Delta_3-\Delta_2)O^{-\epsilon}_{\Delta_3+\Delta_2-1,J=-1}(z_2,\bar{z}_2) +\\
    &+\frac{\bar{z}_{32}}{z_{32}}B(\Delta_3-1,-1-\Delta_3-\Delta_2)G_{\Delta_3+\Delta_2}^{-,-\epsilon}(z_2,\bar{z}_2) \, . \label{eq:G+G-SV}
\end{split}
\end{equation}
While the second line in (\ref{eq:G+G-SV}) agrees with the one in \cite{Pate:2019lpp}, we found the same operator $O^{-\epsilon}_{\Delta_3+\Delta_2-1,J=-1}(z_2,\bar{z}_2)$ as in the OPE (\ref{eq:G+G-}). With the inverted shadow correlator, it should be easier to relate the single-valued correlator to the Mellin transform of a potential graviton amplitude in a non-trivial background in the bulk space-time. We hope to make this precise in future work.

Finally, in the $(13\rightleftharpoons 24)_{\mathfrak{2}}$ channel, the OPE can be obtained from (\ref{eq:G+G-}) by replacing $3$ by $4$ as the single-valued shadow correlator is manifestly symmetric under $3\leftrightarrow4$.

\section{Concluding remarks} \label{sec5}
In the current work, we circumvented the distributional nature of the celestial four-point graviton amplitude by performing a shadow transform on one of the operators. This led to a correlator with non-trivial monodromy that we completed to a single-valued one using the methods in \cite{Fan:2021pbp}. Our graviton correlator shares similar properties with the four-gluon case analysed in \cite{Fan:2021isc,Fan:2021pbp}. The single-valued completion is crossing symmetric, the graviton OPEs from the three channels correspond to the standard celestial ones. Moreover, we can un-shadow the new correlator due to a suitable Coulomb gas representation. One interesting difference is the appearance of an extra spin $J=-1$ field in the leading term OPE in the $(14\rightleftharpoons32)_{\mathfrak{2}}$ and $(13\rightleftharpoons24)_{\mathfrak{2}}$ channels. In our treatment, this is a consequence of crossing symmetry and single-valuedness. However, it suggests that the new correlator might come from gravity coupled to a background field in the bulk. It would be interesting to investigate this further, possibly by leveraging the results of \cite{Ball:2023ukj}.
\par In \cite{Fan:2022vbz}, a single-valued four-point gluon correlator was derived for general values of the conformal dimension by solving the Banerjee-Ghosh equations \cite{Banerjee:2020vnt}. The $\la_1=0$ correlator was used as a boundary condition for the series representation of the more general correlator. Graviton correlators also obey differential equations \cite{Banerjee:2020zlg,Banerjee:2021cly,Ruzziconi:2024zkr}. While more difficult to solve than their gluon counterparts, the double copy clue in the soft limit of one of the operators might help construct a graviton correlator using the known gluon ones. Furthermore, along the lines of \cite{ Stieberger:2022zyk, Stieberger:2023fju}, these new correlators could be embedded in a subclass of Liouville CFT correlators. It would be interesting to make a connection with the recent proposal in \cite{Mol:2024etg,Mol:2024qct}. We leave the exploration of these structures for future work.

The lack of single-valuedness of the shadow correlators is, in our case, a consequence of the distributional nature of the flat space celestial amplitudes. It would be interesting to find a prescription for the shadow transform such that the resultant shadow correlators are automatically single-valued and see the connections with our results. 

It would also be interesting to study multi-shadowed correlators along the lines of \cite{Chang:2022jut, De:2022gjn}. It was shown that the multi-light transformed correlators have the same spectrum as the single-light transformed correlators \cite{De:2022gjn}. We expect this is also case for the multi-shadowed correlators.

Recently, in \cite{Banerjee:2024yir}, the shadow transformation of the conformal primary operators has been shown to be related to the other framework of flat space holography \cite{Jain:2023fxc}. We expect more connections to other approaches to flat space holography (see e.g. \cite{Donnay:2022aba,Bagchi:2022emh,Donnay:2022wvx,Nguyen:2023vfz,Nguyen:2023miw,Mason:2023mti, Liu:2024nfc,Alday:2024yyj,Ruzziconi:2024kzo,Iacobacci:2022yjo,Sleight:2023ojm,Iacobacci:2024nhw,Pacifico:2024dyo}).

\section*{Acknowledgements}

We would like to thank Tim Adamo and Tom Taylor for interesting discussions and valuable comments on the draft. IS is supported by an STFC studentship. BZ is supported by the Royal Society and by the Simons Collaboration on Celestial Holography.

\end{document}